\documentclass[aps,pra,twocolumn,floats,showpacs]{revtex4}
\usepackage{graphicx}
\usepackage{amsmath}
\usepackage{amssymb}
\usepackage{ulem}

\newcommand{\0}{{|0\rangle}}
\newcommand{\n}{{|n\rangle}}

\newcommand{\w}{{\omega}}
\newcommand{\Hmatrix}{{\rm H}}
\newcommand{\Umatrix}{{\rm U}}

\newcommand{\bvec}    {{\bf b}}
\newcommand{\avec}    {{\bf a}}

\newcommand{\W}{{\Omega}}
\newcommand{\Wbar}{\Omega_{\rm eff}}
\newcommand{\Sch}{{Schr\"{o}dinger }}

\newcommand{\Xs} {{X^1\Sigma^+_g}}
\newcommand{\As} {{A^1\Sigma^+_u}}
\renewcommand{\ap}{{\alpha_P}}
\newcommand{\af}{{\alpha_F}}
\newcommand{\R}{{\rm R}}

\newcommand {\ket}[1]{|\,{#1}\,\rangle}

\newcommand{\beq}{\begin{equation}}
\newcommand{\eeq}{\end{equation}}
\newcommand{\bea}{\begin{eqnarray}}
\newcommand{\eea}{\end{eqnarray}}
\newcommand{\eq}[1]{Eq. (\ref{#1})}

\def\mmat#1{\hbox{$\sf{#1}$}}

\def\prl#1#2#3{Phys.~Rev.~Lett.~{\bf #1},\ #2\ (#3)}

\begin{document}
\title{Complete transfer of populations from a single state
to a pre-selected superposition of states using Piecewise Adiabatic Passage.}

\date{\today}
\author{Evgeny A.~Shapiro$^1$, Valery~Milner$^{1,2,3}$, Moshe Shapiro$^{1,2,4}$}

\affiliation{Departments of  Chemistry$^1$ and Physics \&
Astronomy$^2$, and The Laboratory for Advanced Spectroscopy and
Imaging Research (LASIR)$^3$, The University of British Columbia,
Vancouver, Canada, $^4$Department of Chemical Physics, The
Weizmann Institute, Rehovot, 76100, Israel,}

\begin{abstract}{We develop a method for executing robust and
selective transfer of populations between a single level and
pre-selected superpositions of energy eigenstates. Viewed in the
frequency domain, our method amounts to executing a series of
simultaneous adiabatic passages into each component of the target
superposition state. Viewed in {the} time domain, the method works
by accumulating the wavefunction of the target wave packet as it
revisits the Franck Condon region, in what amounts to an extension
of the Piecewise Adiabatic Passage technique [ Shapiro et.al.,
Phys. Rev. Lett. 99, 033002 (2007)] to the multi-state regime. The
viability of the method is verified by performing numerical tests
for the Na$_2$ molecule.}
\end{abstract} \pacs{32.80.Qk,33.80.-b,42.50.Hz}

\maketitle

\section{Introduction}
\label{intro}

The availability of robust  and selective methods of executing
population transfers in multilevel quantum systems is essential
for a variety of fields, such as precision spectroscopy
\cite{Nelson-ChemRev94,Diddams04,DFCS,comb-general,Albert-98},
quantum computing
\cite{QDits1,QDits2,QDits3,QDits4,QDits5,QDits6,QDits7}, control
of molecular dynamics and chemical reactions
\cite{ShapiroBrumerBook,RiceBook,rabitz-science00}, production of
cold molecules \cite{ultracold-review,comb-PA,Shapiro07a},
biophotonics \cite{Malinovskaya07} and nanoscience
\cite{Kral-review,nano}. In this paper we propose a new method for
transferring populations from a single energy eigenstate into a
selected superposition of states (wave packet) using shaped
broadband laser pulses. This simple 
method combines the robustness of adiabatic population passage
\cite{ATheorem-books,AllenEberlyBook,Vitanov01a} with the
flexibility of femtosecond pulse shaping techniques
\cite{ShapiroBrumerBook,RiceBook,Weiner00,Lozovoy-review}.

Our method is an integration of a number of earlier studies.
Viewed in the frequency domain, it is an {application} of
Coherently Controlled Adiabatic Passage (CCAP)
\cite{Kral-review,CCAPs}, which in itself is an extension of the
three-state Stimulated Raman Adiabatic Passage (STIRAP)
\cite{Vitanov01a, STIRAP-OHE, STIRAP-bergmann90} and the adiabatic
transfer between field-dressed states
\cite{Floquet0,Floquet-Guerin,Floquet-Holthaus,Gibson05,Wollenhaupt06a,Baumert-landscapes-08}
methods. The CCAP method presents a complete solution to the
non-degenerate quantum control problem, i.e., the execution of a
complete population transfer between superpositions (wave packets)
of non-degenerate energy eigenstates.

Viewed in the time domain, the present method is an extension of
Piecewise Adiabatic Passage \cite{Shapiro07,PAF,comb-PA} to the
case of a target wave packet of states. In particular, when the
spectrum of the target manifold is nearly harmonic, the driving
field is given by a train of mutually coherent pulses separated by
the {evolution} period of the target wave packet. Keeping the
optical carrier phase of the pulses constant throughout the entire
train of pulses results in the piecewise execution of periodic
(Rabi) oscillations between the initial and the target states. The
introduction of a ``piecewise chirp'', expressed as a
pulse-to-pulse variation in the optical phase, is what eliminates
the oscillations and renders adiabatic robustness to the
population transfer.

Selectivity is obtained by tailoring the temporal and spectral
profiles of the train of pulses to the attributes of the target
wave packet dynamics
\cite{Nelson-ChemRev94,Walmsley-deAraujo,Wollenhaupt06,deAraujo08}.
Such compatibility between the pulse train attributes and the
target wave packet dynamics has also been noted in brute-force
optimization studies
\cite{judson-rabitz,rabitz-science00,Malinovsky97-OCT} aimed at
either maximizing population transfer into the target state
\cite{Grafe05,Trallero-Herrero06,Hertel-JPB08} or stabilizing such
transfer against wave packet spreading and decoherence
\cite{Walmsley-Science08}.

The time profile of the field in our solution is reminiscent of
the ``multi-RAP'' pulse sequences of
Ref.\cite{Baumert-landscapes-08}. As discussed below, the
difference between the two solutions is manifest when the target
wave packet consists of more than two eigenstates. Our method is
also related to the coherent accumulation of transition amplitudes
driven by a train of laser pulses
\cite{Nelson-ChemRev94,DFCS,Diels86,
Temkin93,Vitanov95,Vianna03,RealFields,Jakubetz08}, but the
robustness is a unique property of PAP.

 This paper is organized as
follows: In Section \ref{theory}, we present the theory in the
frequency space and illustrate it via numerical studies of
population transfer in sodium dimers.
 In Section \ref{time}, we describe
the population transfer in the time domain, and connect the two
physical points of view. In Section \ref{SectionNumerics}, we
establish numerically the accuracy of the method and its
sensitivity to the pulse parameters. A discussion is provided in
Section \ref{summary}.

\section{Multi-state adiabatic chirping - frequency domain point of view.}
\label{theory} In this section we extend the CCAP
\cite{Kral-review,CCAPs} method of using a multi-mode pulse to
execute wave packet adiabatic passage via an intermediate state to
the chirped multi-mode pulse case. As in
Refs.\cite{Kral-review,CCAPs}, we consider a quantum system,
initially in the ground state $\0$, coupled by a laser field field
$\epsilon(t)$, made up of $N$ discrete modes, of frequencies
$\omega_n$, to a manifold composed of $N$ excited states. Contrary
to Refs. \cite{Kral-review,CCAPs} we allow the mode frequencies to
slowly vary with time, hence we denote them as $\omega_n(t)$.

The material Hamiltonian is 
%
\beq
  \hat H = \hat H_M + \hat\mu\epsilon(t)
= \hat H_M + 2 f(t) \hat\mu\sum_{n=1}^N\epsilon_{n}
\cos[\Phi_n(t)+\phi_{n}], \label{hamiltonian} \eeq where $\hat
H_M$ is the field-free Hamiltonian, $2f(t)\epsilon_{n}$ and
$\phi_{n}$ are the mode amplitudes and phases, respectively.
$\hat\mu$ is the dipole moment for the transition between the
ground state and the manifold of excited states. We assume that
each mode frequency $\omega_n(t)=\dot\Phi_n(t)$ is detuned by a
small amount $\Delta_n(t) \equiv E_n-E_0-\w_n(t)$ with respect to
one of the excited levels (denoted $n$), where $E_0$ and $E_n$ are
the field-free energies of the states $\0$ and $\n$, respectively,
\begin{equation}
(E_0-\hat H_M)|0\rangle = (E_n-\hat H_M)|n\rangle = 0 \ .%
\end{equation}
The material wave function expressed in a.u. ($\hbar=1$) is expanded as
\begin{equation}\label{bigexpansion-manychannels}
|\Psi(t)\rangle = b_0(t) e^{-iE_0t}|0\rangle
     + \sum_{n=1}^Nb_n(t) e^{-i(E_0 t+\Phi_n(t))}|n\rangle \ .
 \end{equation}
It is instructive to study the dynamics arising from the
application of the Rotating Wave Approximation (RWA)
\cite{AllenEberlyBook,ShapiroBrumerBook,RiceBook,Vitanov01a}. To
do so we first average the non-stationary \Sch equation over time
scales of the order of $1/\w_n,$ resulting in
\begin{eqnarray}\label{RWAed-1}
i \, \dot b_0(t) &=&%
f(t) \sum_{n=1}^N  b_n(t) \mu_{0n}
\sum_{l=1}^N\epsilon_l e^{i((\w_l - \w_n) t + \phi_l)}~,
\\
i \, \dot b_n(t) &=& \Delta(t) b_n(t)
+ f(t) b_0(t) \mu_{n0}\sum_{l=1}^N\epsilon_l
e^{i((\w_n - \w_l) t - \phi_l)}.
\nonumber
\end{eqnarray}
We also assume
that each transition is driven by only one of the field modes -
the mode with the smallest $\Delta_n(t)$ detuning.
Denoting by $\W_n(t)\equiv f(t)\epsilon_n\mu_{n0}\,e^{-i\,\phi_n}$
the (complex) Rabi frequencies, and by $\bvec(t) =
\left(b_0(t),b_1(t)...b_n(t)\right)^{\mmat T},$ the (column)
vector of expansion coefficients,
with the superscript ${\mmat T}$ marking
the ``transpose'' operation, we can now write
the \Sch equation in matrix form as
\begin{equation}\label{MatrixSch}
i\,\dot\bvec = \Hmatrix \bvec \ ,
\end{equation}
where
\begin{equation}\label{Hmatrix}
\Hmatrix(t) = \left(%
\begin{array}{cccc}
  0         & \W_1^*(t)   & \ldots  & \W_N^*(t) \\
  \W_1(t) & \Delta_1(t) & \ldots  & 0 \\
  \vdots    & \vdots & \ddots  & \vdots  \\
  \W_N(t) & 0      & \ldots  & \Delta_N(t) \\
\end{array}%
\right) \ .
\end{equation}
While the first part of the RWA  (Eq.(\ref{RWAed-1})) amounts to
neglecting the terms which oscillate at optical frequencies, the
second part (Eqs.(\ref{MatrixSch},\ref{Hmatrix})) is equivalent to
averaging the \Sch equation over time scales of the order of
$1/(\w_n-\w_m)$, resulting in loss of information about times
scales shorter than a vibrational period. Nevertheless, the
time-averaged \Sch equation (\ref{MatrixSch}) provides an accurate
description of the wave function at the end of the process, after
many vibrational periods. We postpone the study of the dynamics on
a finer time scale to the following sections.

We now tune the chirping of the mode frequencies such that all the
detunings are equal to one another, i.e., $\Delta_n(t)=\Delta(t).$
This allows us to easily diagonalize $\Hmatrix$ at any given
moment of time, obtaining $N-1$ degenerate ``dark'' eigenstates
whose (quasi-) energies are equal to $\Delta(t)$, and two
``bright'' eigenstates whose eigenvalues are,
\begin{equation}\label{lambdapm}
\lambda_\pm(t)
 = \Delta(t)/2\pm \{\Delta^2(t)/4+ \Wbar^2(t)\}^{1\over 2}
\end{equation}
where $\Wbar^2(t)= \sum_{n=1}^N |\W_n(t)|^2.$
The eigenvectors corresponding to the bright solutions are
\begin{equation}\label{b_pm}
\bvec_\pm = \cos\theta_\pm \bvec_i %
 + \sin\theta_\pm \bvec_f
\ ,
\end{equation}
where
\begin{equation}
\bvec_i= (1,0,\ldots,0)^{\mmat T}~~~
\bvec_f= (0,\W_1,\ldots,\W_N)^{\mmat T}/\Wbar \ , \label{Bvec}
\end{equation}
with~ $\tan\theta_\pm=\lambda_\pm/\Wbar.$
Note that each $\W_n$, as well as $\Wbar$,
depends on time via the common factor $f(t)$.
Therefore the definition of $\bvec_f$ in Eq.(\ref{Bvec})
is time independent. Further,
$\tan\theta_+\cdot\tan\theta_-=-1,$ and $\lambda_-<0$, hence,
\begin{equation}\label{sinpm}
\sin\theta_+=\cos\theta_-, \qquad \cos\theta_+ = -\sin\theta_- \,.
\end{equation}
If at some instant $|\Delta(t)|\gg \Wbar(t),$ then one of the states represented by
the $\bvec_+$ or $\bvec_-$ vector coincides with the $\ket{i}$ state,
represented by the $\bvec_i$ vector, and the other - by the $\ket{f}$ state,
represented by the $\bvec_f$ vector.

\begin{figure} \centering
\hskip -.65truein
 \includegraphics[width=1.1\columnwidth]{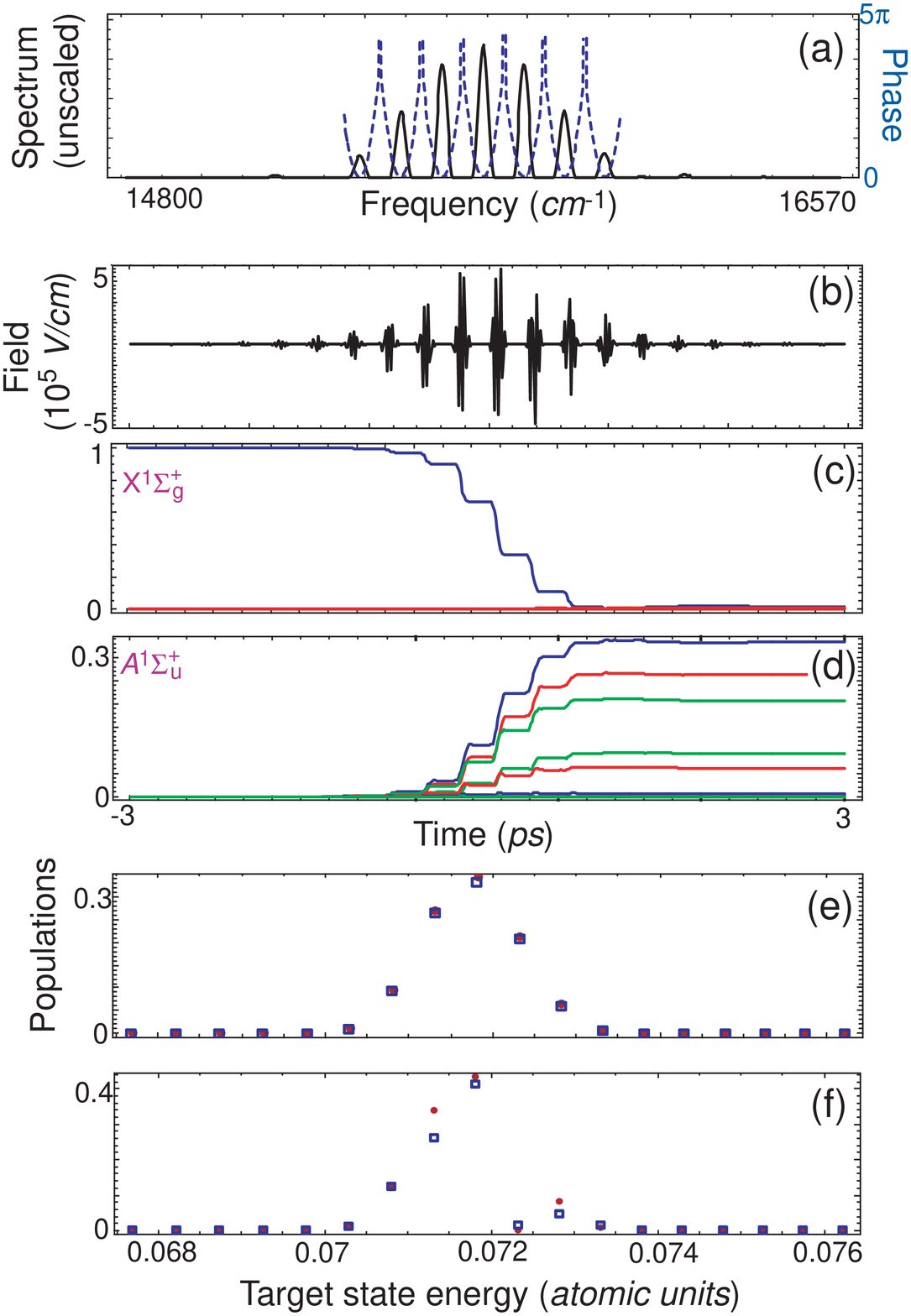}
\caption{(Color online) Piecewise adiabatic following in Na$_2$.
(a): Amplitude spectrum (solid black) and frequency-dependent
phase (dashed blue) of the driving field. (b): The driving field
as a function of time. (c,d): Populations of various vibrational
levels of the ground and excited electronic states. (e): Final
populations of $\As$ states predicted by the theory (full red
circles), and obtained in the numerical simulation (empty blue
squares).(f): Same as (e) for the case when the spectral region
corresponding to the transition into one of the excited levels is
blocked.}
       \label{FigPAF}
\end{figure}

Similar to the textbook case of adiabatic following in a two-level
system \cite{AllenEberlyBook}, when $\Delta(t)$ is made to vary slowly enough,
the system makes a smooth transition from state $\ket{i}$ to
state $\ket{f}.$
We show in the Appendix that there are no transitions between the
$\ket{\pm}$ bright eigenstates and the dark
states. As a result, the dark states remain unpopulated at all
times. Hence it is possible to make the adiabatic population transfer
complete, provided that
the bandwidth of the pulse covers the whole range of the target
energies $E_n$.

In order to create the required field, one can start with a single
broadband laser pulse, and spectrally shape it in the way
illustrated in Fig.~\ref{FigPAF}(a). Here, the field is frequency
chirped in the neighborhood of each resonance frequency
$\omega_n^{res}= (E_n-E_0)$, according to,
\begin{equation}\label{manychirps}
\epsilon(\omega) = \epsilon_n\, F(\w-\w_n^{res})\,
e^{i\,{\alpha_\omega} (\omega-\omega_n^{res})^2/2} \,e^{i(\w
t_0-\phi_n)} \ .
\end{equation}
 The real amplitude
envelope $F(\w)$ reaches its maximum of 1 at $\omega =0$ and
serves to suppress the pulse spectrum between the resonance
regions. The requirements for this suppression will be discussed
in Section \ref{SectionNumerics}.
For Gaussian profile of the field envelope near each
$\omega_n^{res}$,
\begin{equation}
F(\omega-\w_n^{res})=
\exp\left[(\omega-\omega_n^{res})^2/2\sigma_\omega^2\right]
\label{GaussianModulation} ~.%
\end{equation}
The above form corresponds to the field parameters of \eq{hamiltonian}
assuming the form,
%
%
\begin{eqnarray}
\Phi_n(t) &=&\w_n^{res} (t-t_0) +\frac{\alpha_t}{2}
(t-t_0)^2+\phi_c  ,
\nonumber\\
f(t) &=& {\sigma_w (2\pi)^{1\over 2}\over
(1+\alpha_\w^2\sigma_\w^4)^{1\over 4}}\exp\left[-(t-t_0)^2/2\sigma_t^2\right],
\nonumber\\
\nonumber\\
\phi_c &=& -{\rm arg}\left[1-i\alpha_\w\sigma_\w^2\right]/2  ,
\label{alphat}
\\
\nonumber\\
 \sigma_t^2 &=&
(1+\alpha_\omega^2\sigma_\omega^4)/\sigma_\omega^2 ~,\nonumber\\
\alpha_t &=& \alpha_\omega \sigma_\omega^4/
(1+\alpha_\omega^2 \sigma_\omega^4) \ . %
\nonumber
\end{eqnarray}
In the adiabatic transfer into a pre-selected superposition state,
the real amplitudes $\epsilon_n=\epsilon(\omega_n^{res})$ and
phases $\phi_n$ are chosen such that the vector $\bvec_f$ given by
Equation (\ref{Bvec}) represents the $\ket{f}$ target state. The
direct correspondence between Eqs.(\ref{lambdapm},\ref{b_pm}) and
the equations describing adiabatic following in a two-level system
driven by a single-component chirped pulse \cite{AllenEberlyBook}
suggests that the variation of each chirped resonant frequency
must exceed $\Wbar$.

Fig.~\ref{FigPAF} shows a simulation of the adiabatic transfer
between the $v=0$ vibrational state of the Na$_2(\Xs)$ ground
electronic state, and a vibrational wave packet composed of the
$v=7,..,12$ states of the $\As$ state \cite{Na2}. In order to
assess the fidelity of the transfer we have also computed the
population of neighboring $\Xs$ and $\As$ vibrational levels.
Panel (a) shows a field spectrum obtained by frequency modulating
a $\sin^2\beta t$-type pulse. The pulse duration is 55fs and its
central wavelength is 638 nm, with chirp parameter $\alpha_\omega
= 2\times 10^{5}$fs$^2$. As discussed in detail in the next
section, the piecewise chirped pulse of panel (a) corresponds in
the time domain to the train of pulses shown in
Fig.~\ref{FigPAF}(b).

In Fig.~\ref{FigPAF}(c-e) we display the dynamics of the
population transfer. As demonstrated in Fig.~\ref{FigPAF}(e), the
populations of the target wave packet energy components
practically coincide with the predictions of Eq.(\ref{Bvec}). In
Fig.~\ref{FigPAF}(f) we demonstrate the selectivity of the method
by changing the amplitudes $\epsilon_n$ and the
 phases $\phi_n$ characterizing the field components. In each case,
the distribution of the final populations in the target manifold
remains close to that predicted by Eq.(\ref{Bvec}), though
the component phases could at times deviate
from the target ones,
somewhat lowering the magnitude of the overlap
with the target wave packet to $\sim
0.8-0.95$. This point is discussed further
in section \ref{SectionNumerics}.

\begin{figure}[b]
\centering
\hskip -.29truein
 \includegraphics[width=1.05\columnwidth]{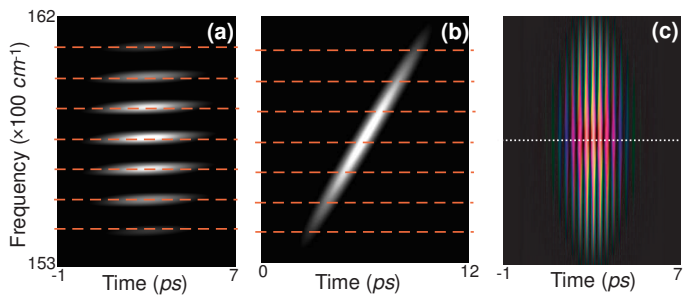}
\caption{(Color online) Field spectrograms. (a): Absolute values
of the overlap integral of the pulse train field with probe
Gaussian pulses. The red horizontal lines show the resonance
frequencies $\w_n^{res}$. (b): Spectrogram of a conventional
frequency-chirped pulse.  (c): Expansion of the pulse train field
in short probe pulses. The amplitude of each projection is shown
by brightness, its phase by color. The white dotted line marks the
spectral center of the pulse. }
       \label{FigHusimi}
\end{figure}

The mechanism of the population transfer becomes transparent if
the laser field is represented by a phase-space (Husimi)
spectrogram. Fig.~\ref{FigHusimi}(a) shows the spectrogram of the
field of Fig.~\ref{FigPAF}(a,b). Each point displays the absolute
value of the overlap integral of the field with a Gaussian probe
pulse,
\begin{equation}
\label{probe} \epsilon^{(p)}(t) = \exp\left[-{1\over 2}
\left({t-t_0^{(p)}}\over \sigma_t^{(p)}\right)^2
-i\w^{(p)}t+i(\w^{(p)}-\w_0) t_0^{(p)}\right]
\end{equation}
of long duration ($\sigma_t^{(p)}=1000$ fs). The frequency and time axes
correspond to $\w^{(p)}$ and $t_0^{(p)}$, respectively. It is
clear that the field drives a number of adiabatic passages
simultaneously: An adiabatic transition into
$n^{\rm th}$ excited level
may occur when the instantaneous frequency component $\w_n$
crosses the $n^{\rm th}$  resonance at $\w_{n}^{res}=E_n-E_0$.

The mechanism displayed here is quite different from that of the
``molecular $\pi$-pulse''
\cite{Cao98,PiPulses-exps1,PiPulses-exps2,PiPulses-exps3,Malinovsky01},
where a transition into the manifold of excited states is driven
by a single frequency-chirped laser field. While in the scheme
proposed here all the transitions take place simultaneously, in
the case of a molecular $\pi$-pulse the energy of the initial
state shifted by the energy of  one photon, [$E_0 + \w(t)$],
crosses all the target energies one by one. As a result, the
system passes through a chain of avoided crossings. This is
illustrated in Fig.~\ref{FigHusimi}(b) which shows the spectrogram
($\sigma_t^{(p)}=200$fs) of the molecular $\pi$-pulse obtained
from the original un-shaped field, used in our numerical examples,
by applying a single frequency chirp of $\alpha_\w=2 \times
10^{4}$ fs$^2$. Although a molecular $\pi$-pulse can transfer all
the initial population into the excited manifold in a robust way,
the relative amplitudes of the states in the resulting wave packet
are not controllable unless the target manifold consists of only
two states. In the latter case, employing either up- or down-
frequency chirp allows population transfer in either the lower or
the higher eigenstate of the target manifold
\cite{Wollenhaupt06a}. We refer the reader to
Ref.\cite{Malinovsky01} for a detailed analysis of the population
transfer by molecular $\pi$-pulses.

The question of selectivity also arises when comparing our method
to the family of Stark-assisted adiabatic transfers
\cite{Sola-StarkSelective-PRA07}, and the transfer via multiple
successive Rapid Adiabatic Passages \cite{Baumert-landscapes-08}.
While in both examples the possibility of robust and selective
transfer into either one of the two target states has been found,
it is not clear whether these methods can enable population
transfer into a selected superposition of target states,
especially if $N>2$.

\section{Multi-state adiabatic chirping - temporal point of view.}
\label{time}

In order to understand the dynamics of the population transfer on
a finer time scale, here we develop the time-domain description of
the process. Fig. \ref{FigPAF}(b) shows the time-dependent driving
field of the above example. The time analogue of
Eq.(\ref{manychirps}) is
\beq \label{field-timerep} \epsilon(t) = R_e \, f(t) \exp
\left[i\omega_0 (t-t_0)+i\alpha_t(t-t_0)^2/2+i\phi_c \right] g_1
(t)  \, \eeq
where  $\omega_0 = \w_{n_0}^{res}$ is chosen to coincide with the
frequency of a transition into one of the central states
$|n_0\rangle$ of the excited wave packet, and
\begin{equation}\label{G1}
g_1(t) = 2\sum_{n=1}^{N} \epsilon_{n}
\exp\left[i(E_n-E_{n_0})(t-t_0)+i\phi_n\right] \ .
\end{equation}
Equation (\ref{G1}) is similar to an expression describing wave
function dynamics of a fictitious wave packet in the target
manifold, with the amplitudes of eigenstates given by
$\epsilon_n\exp[i\phi_n]$. If the target spectrum and the
distribution of the complex field amplitudes
$\epsilon_n\exp[i\phi_n]$ are both smooth, then the interference
between the different frequency components of the field results in
a train of short pulses separated by the vibrational period
$T_{vib} = dE_n/dn$  of the target wave packet. The field pattern
is restored whenever $(E_n-E_{n_0})t\simeq 2\pi m$ for all $n$ and
any integer $m$. Such a train of short pulses is seen in
Fig.~\ref{FigPAF}(b). Further, if the target spectrum is harmonic,
then the field spectrum is a frequency comb with equally spaced
teeth, and the pulse shape is preserved within the train. If the
spectrum is anharmonic, then the shape evolves from pulse to pulse
within the train, reflecting the spreading of the target wave
packet as it revisits the transition region.


When the target spectrum is weakly anharmonic, the pulse train of
Eq. (\ref{field-timerep}) is composed of pulses which differ from
one another in four essential ways: First, the temporal spacings
between the adjacent pulses (which may have different overall
power) change. Second, their central frequencies differ by
$\alpha_t\, T_{vib}$. Third, and most important, the $l^{\rm th}$
pulse has an extra overall phase equal to $\alpha_t\, (lT_{vib})^2
/2$, where $l=0$ corresponds to the middle of the pulse train; and
fourth, owing to the anharmonicity, the condition $(E_n-E_{n_0})t
= 2\pi m$ is fulfilled for different $n$ at slightly different
times, leading to change in the pulse duration from pulse to
pulse.

Fig.~\ref{FigHusimi}(c) shows the spectrogram of the field of
Fig.~\ref{FigPAF}(a,b) using short ($\sigma_t^{(p)}=50$fs) probe
pulses, whose exact form is given by Eq. (\ref{probe}). Naturally,
the representation in terms of short probe pulses gains temporal
resolution but loses frequency resolution. The optical phase of
the short pulses within the train varies quadratically with the
pulse number. This is displayed in Fig. \ref{FigHusimi}(c) by the
variation in color along the central-frequency section, from
orange (phase equal to zero) in the middle of the train, to purple
and blue, and finally to green (phase $\simeq 3\pi/2$), at the
very tails. Note that the pulse-to-pulse drift of the central
frequency is not noticeable on the scale of the figure.

\begin{figure}[b]
\centering
 \includegraphics[width=0.7\columnwidth]{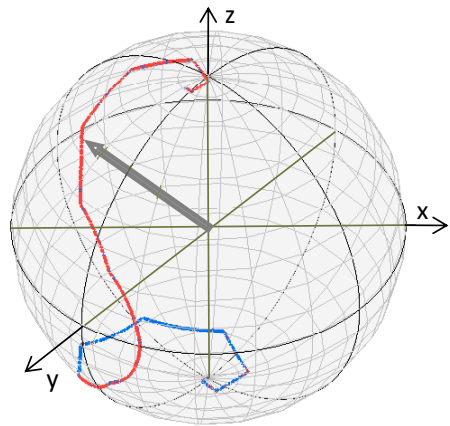}
\caption{(Color online). A calculated sample trajectory of the
Bloch vector (thick gray arrow) during the a piecewise AP process
implemented with a train of 20 ultra-short pulses \cite{PAF}.}
  \vskip -.1truein
\label{FigBloch}
\end{figure}

The field in the above example is similar to that introduced in
Ref.\cite{PAF} for the piecewise adiabatic passage in a two-level
system. This is readily seen both in the frequency-domain
representation (by comparing Fig.\ref{FigPAF}(a) with Fig.3(b) of
Ref.\cite{PAF}), and in the time-domain representation. The
almost-periodic pulse train only drives transitions between the
$\ket{i}$ and $\ket{f}$ states. The train of short pulses,
tailored to fit the periodic evolution of the $\ket{f}$ state,
peaks every time this state revisits the Franck-Condon region. At
these instants a  superposition state $\ket{\psi}$ can be
represented by the unit Bloch vector of Fig.~\ref{FigBloch},
defined by the $\theta$ and $\phi$ angles,
\begin{equation}
\ket{\psi}= \cos(\theta/2)  |i\rangle +
      \sin(\theta/2) e^{i\phi} |f\rangle~, \label{twostate}
\end{equation}

The present 1+many-level system differs from the two level system
in a fundamental way, since here the relative phases of the
components of the target wave packet change all the time. There
are points of similarity as well, because here too, the process
can be depicted, as shown in Fig. 3, by the discontinuous motion
of a unit vector on a (``Bloch'') sphere, with the polar angle
$\theta=0$ representing $\ket{i}$ and $\theta=\pi$ representing
$|f\rangle$. The effect of each pulse (of short duration $\tau$)
is now viewed as a rotation ${\hat P}\equiv\R(\ap)$ of the Bloch
vector by an angle $\ap=\int_{\tau} \Wbar (t) dt $ about the $y$
axis. The change in the carrier phase between consecutive pulses
can be represented by an additional rotation about the $z$ axis,
${\hat F}\equiv\R_z(\af).$ Thus, the overall evolution due to the
pulse train of Eq. (\ref{field-timerep}) is represented by a
sequence $\hat U = ...\hat T \hat F \hat P \hat T \hat F \hat P
...,$. The operator $\hat T,$ which corresponds to the free
evolution of the wave packet between pulses, keeps the Bloch
vector orientation at the beginning of each short pulse in the
train equal to that at the end of the previous one.

The product $\hat F \hat P$ of two rotations
can be viewed as an overall rotation by an angle $\alpha_0$
about an instantaneous axis defined by the ($\theta_0,\phi_0$) angles, given to
lowest-order expansion in $\ap$, $\af$ as

$$
\alpha_0= \{(\ap^2+\af^2)/2\}^{1\over 2},~~~~~~~
\phi_0=  \pm\pi/2-\af/2,~~~
$$
\beq
\tan\theta_0 =  \pm \ap/\af. \label{a0}
\eeq
By maintaining the same value of $\af$ and $\ap$ throughout the pulse train
we can induce piecewise rotations of the Bloch vector about the
``adiabatic'' trajectory traced out by
$(\theta_0,\phi_0)$. As shown in Fig. 3, by slowly varying the values
of $\ap,\af$ we can make the Bloch vector follow this adiabatic trajectory.
Such piecewise following can be realized provided,
(i) the $y$- and $z$-rotations are small (i.e. each pulse should induce an
angular change much smaller than $\pi$ and each increment in the carrier
phase should be small too);  (ii) ($\theta_0,\phi_0$) should not move much
from pulse to pulse, i.e.
\begin{equation}\label{ClassicalAdiabaticity}
\Delta\theta_0 \ll \{(\ap^2+\af^2)/2\}^{1\over 2}.
\end{equation}

In the pulse train of Eq. (\ref{field-timerep}),
initially $\ap\ll |\af|$: $\Wbar$ is small, while the pulse-to-pulse phase
change is significant.
As $\ap$ increases and $|\af|$ decreases, the states originating in
$|i\rangle$ and $|f\rangle$ move towards the equator of the Bloch sphere.
They cross the equator as soon as $\af$ changes sign, and finally interchange
with each other. Depicted in the original non-rotated frame, the trajectory
of the Bloch vector is a piecewise spiral, similar to that shown in
Fig.\ref{FigBloch}.

\section{Numerical studies.}
\label{SectionNumerics}

\begin{figure} \centering
 \includegraphics[width=0.99\columnwidth]{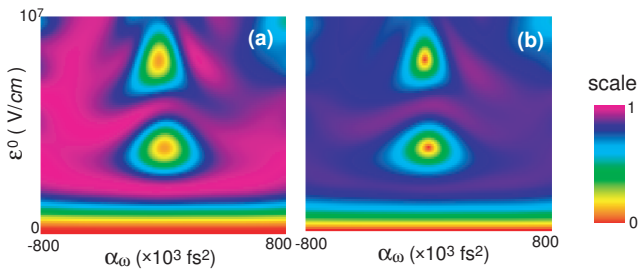}
\caption{(Color online) Population transferred into the target
manifold (a), and the projection on the target wave packet (b) as
a function of the chirp $\alpha_\omega$ and field strength
$\epsilon^{0}$.}
       \label{FigScan}
\end{figure}

Fig.~\ref{FigScan} shows the population transferred into the
target manifold, and the projection of the final state onto the
target wave packet comprised of vibrational eigenstates of Na$_2$
in the $\As$ electronic state as a function of the chirp
$\alpha_\omega$ and the field strength $\epsilon^0$ of the
original 55fs pulse used in the computations presented in
Fig.~\ref{FigPAF}. The plots reveal several interesting features.
First, for both positive and negative chirps exceeding in
magnitude $|\alpha_\omega|\sim 200000$fs$^2$, the transfer is
almost complete and quite robust with respect to changing
$\alpha_\omega$ and $\epsilon^0$. For a large range of pulse
parameters the transfer probability is $\sim95\%$, and the
projection of the final state onto the target is $\sim 0.85-0.9$.

The transfer probability landscape at small values of
$\alpha_\omega$ corresponds to the piecewise Rabi oscillations
between the initial state and the excited wave packet. A close
look at the population dynamics in this region of parameter space
shows that the first few pulses in the train manage to completely
empty state $|0\rangle$, while the following pulses re-populate
it. At stronger fields, state $|0\rangle$ is de-populated and
re-populated several times during the pulse sequence. Populations
of the nearby vibrational eigenstates of the $\Xs$ manifold remain
negligible at all times, with the ratios of populations of
different $\As$ states remaining close to the target values.

We also note that though the parameter-space region where robust
population transfer occurs includes pulse areas of $\sim\pi$, the
optimal pulses have areas larger than $\pi.$  The pulse area
cannot be increased beyond a certain value because when individual
pulses within the pulse train are able to drive a significant
population transfer, condition (i) above breaks down, and the
overall fidelity decreases.

The fidelity profile of Fig.~\ref{FigScan} is quite robust with respect to
changes in the spectral width $\sigma_\w$
in the vicinity of each resonance frequency $\omega_{n}^{res}$.
However, when $\sigma_\w$ approaches half of the distance between
the neighboring resonance frequencies, the transfer fidelity
exhibits ripples as a function of $\alpha_\w$, eventually becoming
unstable. Indeed, for the Gaussian amplitude modulation of the
spectrum (Eq.(\ref{GaussianModulation}-\ref{alphat})), and
assuming that the chirp is strong so that
$\alpha_\w\sigma_\w^2\gg1$, one has
\begin{equation}
\sigma_t = \alpha_\w \sigma_\w, \qquad \alpha_t =
1/\alpha_\w\ ,
\end{equation}
and the instantaneous frequency of the $n^{\rm th}$ component of the
field at the end of the pulse train becomes
\begin{equation}
\w_n(t=t_0+2\sigma_t)~ \w_n^{res}+2\sigma_t \alpha_t \simeq
\w_n^{res}+2\sigma_w \ .
 \label{FinalW}\end{equation}
If $\sigma_\w \geq (\w_{n+1}^{res}-\w_n^{res})/2$ then the
instantaneous frequency of the $n^{\rm th}$ field component at the end
of the pulse train coincides with the adjacent
resonance. In this case the dynamics can not be viewed as a set of
independent parallel adiabatic passages into different target
states.

\section{Summary and discussion.}
\label{summary}

In this paper we have developed a method for executing robust and
selective transfer of population from a single energy eigenstate
to a pre-selected superposition of energy eigenstates. Viewed in
the frequency domain, the method constitutes simultaneous transfer
of population to all the energy eigenstates which make up the
superposition state by a set of parallel adiabatic passages.
Viewed from the time domain, the method amounts to using a train
of pulses to accumulate wavelets which make up the target wave
packet as it revisits the Franck Condon region
\cite{Shapiro07,PAF, Shapiro07a}.

We have tested the method numerically by simulating
transitions between a single vibrational eigenstate and a
superposition of vibrational energy eigenstates of Na$_2$ in the $\As$ state.
The simulations confirmed
the high selectivity and robustness of the method.

Topics to be investigated further include: An accurate description
of the transfer dynamics for a general anharmonic spectrum and in
particular the description of the short-time dynamics of the
transfer. This may be achieved using Floquet states
\cite{Floquet-Guerin,Floquet-Holthaus} dressed by a multi-mode
driving field. Also, although the present theory is quite accurate
in predicting the ratio of populations transferred into the target
wave packet, the reasons for the phase errors, resulting in less
than perfect overlap with the target state, need to be
investigated.


\begin{acknowledgements}
This work was supported by grants from CFI, BCKDF and NSERC.
\end{acknowledgements}

\section{Appendix: Proof that the dark states remain unpopulated.}

This proof follows Ref.\cite{Kral-review}, where
similar arguments were made in the context of CCAP. These arguments were
based on
the standard treatment of adiabatic
transfer \cite{ATheorem-books,AllenEberlyBook,ShapiroBrumerBook}.

It follows from Eq.(\ref{Hmatrix}) that each dark state is
described at a given time by the vector of amplitudes
$\bvec_{dark}=(0,b_{1\bot},...,b_{N \bot} )^{\mmat T}$ which is orthogonal
to both $\bvec_i=(1,0,...0)^{\mmat T}$ and
$\bvec_f=(0,\W_1/\Wbar,...,\W_N/\Wbar)^{\mmat T}$. Since the mutual ratio
of different Rabi frequencies $\W_n$ does not change in time, we
can choose for the dark states an $N-1$-dimensional
time-independent basis $(0,e_{1\bot}^{(n)},...,e_{N\bot}^{(n)})^{\mmat T},
\ n=1..N-1$. Introducing the vector $\avec$ of amplitudes of
instantaneous eigenstates,
\begin{equation}\label{avec}
\bvec(t) = \Umatrix(t)\avec(t) \ ,\end{equation}
where the columns of $\Umatrix$ are given by the normalized
eigenvectors of $\Hmatrix$, and using the relation
$\Umatrix^\dag\Umatrix=1$, we have that
\begin{equation}
\label{avec-evolution} %
i\, \dot\avec = \Umatrix^\dag \Hmatrix\Umatrix \,\avec -
i\,\Umatrix^+\dot\Umatrix \,\avec \ .
\end{equation}
The first term on the right-hand side of Eq.(\ref{avec-evolution})
governs the adiabatic evolution. The second term governs
non-adiabatic transitions between the instantaneous eigenstates.
Let us choose the order of adiabatic eigenvalues such that the
adiabatic Hamiltonian $\Umatrix^\dag \Hmatrix\Umatrix$ has on its
main diagonal $(\lambda_+,\lambda_-,\Delta,...\Delta)$, with $N-1$
terms equal to $\Delta$, and all the off-diagonal elements equal
to zero. Then, using Eq.(\ref{sinpm}), we obtain
\begin{widetext}
\begin{equation}
\label{U}
\Umatrix = \left(%
\begin{array}{ccccc}
  \cos\theta_+                   &   \sin\theta_+                  & 0                &\ldots  & 0 \\
  \sin\,\theta_+\,{\W_1}/{\Wbar} & -\cos\theta_+\,{\W_1}/{\Wbar}   & e_{1\bot}^{(1)}  &\ldots  & e_{1\bot}^{(N-1)} \\
  \vdots                         & \vdots                          & \vdots           &\ddots  & \vdots  \\
 \sin\theta_+\,{\W_N}/{\Wbar}    & -\cos\theta_+\,{\W_N}/{\Wbar}   &~~e_{N\bot}^{(1)} &\ldots  & ~~e_{N\bot}^{(N-1)}. \\
 \\
\end{array}%
\right) ~~~~
\end{equation}
\end{widetext}
As a result of the above structure of $\Umatrix,$ the
non-adiabatic matrix is given as,
\begin{equation}
\label{UUdot}
\Umatrix^\dag\dot\Umatrix = \left(%
\begin{array}{cc}
\begin{array}{cc}
0                        & \dot\theta   \\
-\dot\theta         &0                  \\
\end{array}
& \bf 0 \\
\bf 0        & \bf 0  \\
\end{array}%
\right) \ .
\end{equation}

We see that this matrix does not couple the bright and dark
states. Hence the $N-1$ dark states remain unpopulated at all
times.


\end{document}